# AN INVESTIGATION OF THE MONITORING ACTIVITY IN SELF ADAPTIVE SYSTEMS

Yousef Abuseta

Computer Science Department, Faculty of Science
Al-Jabal Al-Gharbi University
Zintan, Libya

## ABSTRACT

*Runtime monitoring is essential for the violation detection during the underlying software system execution. In this paper, an investigation of the monitoring activity of MAPE-K control loop is performed which aims at exploring:(1) the architecture of the monitoring activity in terms of the involved components and control and data flow between them; (2) the standard interface of the monitoring component with other MAPE-K components; (3) the adaptive monitoring and its importance to the monitoring overhead issue; and (4) the monitoring mode and its relevance to some specific situations and systems. This paper also presented a Java framework for the monitoring process for self adaptive systems.*

## 1. INTRODUCTION

A self-adaptive system (SAS) is a system that is able to change its structure or behavior at run-time in response to the execution context variations and according to adaptation engine decisions [1, 2]. SASs have emerged to overcome and handle the complexity of today's software that stems from the dynamic nature of the operational conditions in which these systems have to operate. Such conditions include unstable resource availability, existence of errors that are hard to predict and changing user requirements [3].

A considerable number of research papers has been published in which many approaches and paradigms have been proposed to design and develop self adaptive systems. Such approaches can be classified into many different areas of software engineering such as requirements engineering [4], design patterns [5, 6] software architecture [7, 8, 9] and component-based development [10]. Each approach has tackled the problem from a different aspect and suggested its solution accordingly. However, most of these approaches are centered around the idea of the feedback control loop. One well known example of this kind of loops is the IBM blueprint [11] where the loop consists of four activities, in addition to the Knowledge base, that are responsible for the fundamental functions of self-adaptation. Such activities referred to as *Monitor*, *Analyze*, *Plan*, and *Execute*.

The monitor activity plays a central role in the feedback control loop since it is the first to take place in the loop and thus subsequent activities of the loop depend on its degree of accuracy. Despite that, there is still no a standard framework for the monitoring activity which accommodates the necessary reusable process elements and system components that enable modeling and engineering the monitoring in an accurate, extensible and adaptive manner.

  33

In this paper, we investigate the monitoring activity of the MAPE-K control loop in terms of the following aspects:

- The architecture of the monitoring activity which includes the fundamental components and the control and data flow between these components.
- The monitoring modes and their relevance to some specific situations.
- The appropriate and standard monitor interface with other activities of MAPE-K control loop (the analysis and Knowledge base).
- The adaptive monitoring and its importance to realizing cost effective (in terms of time, resource utilization, etc) self adaptive systems.

We also propose a Java framework that contains the necessary programs (classes, interfaces,..etc) that enable the provisioning of the monitoring activity in a standard, extendible and reusable way. The rest of the paper is organized as follows. Section 2 serves as a background on related concepts to the work of this paper. In section 3, an analysis of the construction of the monitoring model is introduced. Section 4 introduces the process of engineering the functionality of the core application. Section 5 presents a Java implementation of the proposed framework. Section 6 reviews some related work on the monitoring activity design and implementation The paper is concluded in section 7 with some directions for future work.

## 2. BACKGROUND

### 2.1 Monitoring Activity

Monitoring aims at collecting relevant data from one or more components of the software system and its environment in order to keep the system state and behavior at a desirable range that complies with the system goals. In [11], the monitoring is defined as a capability that provides an extensible run-time environment for an autonomic manager to gather and filter data obtained through sensors

The monitoring activity in self adaptive systems is achieved through the deployment of necessary infrastructure, such as sensors and databases. It is based on collecting relevant data of certain attributes that reflect the system internal state (e.g., available resources, errors and faults) and the environment (e.g., user's device and user's location). However, the cost of monitoring is tend to be very high both in terms of resources and processing time. The deployment of a comprehensive monitoring infrastructure can be as expensive as the implementation of the monitored core application. This is especially true when considering online monitoring in self-adaptive systems, where the monitoring process has to continuously collect and process information at runtime [12]. The more data and more frequently is collected, the higher the cost is. This cost takes the form of system slowness or data-management overhead. Therefore, to overcome the issue of monitoring overhead, the monitoring system must behave in an adaptive manner. The adaptation of monitoring may be achieved by either adjusting at runtime the set of metrics to be monitored or decreasing the measurement frequency to reduce the overhead.

### 2.2 Principle of Software Change

Self adaptation is highly connected with performing changes to software systems that need to be made self adaptive. Thus, the change action plays a crucial role in the success of designing such systems. In [13], a taxonomy of software change were presented where the primary aim was to classify different tools and approaches with regard to the software evolution domain. Many aspects related to software change are of a great importance and affect directly the construction of the monitoring model. Such aspects were classified into four categories, namely *the temporal properties*, *object of change*, *system properties* and *change support*. However, we will concentrate here only on the aspects that we believe they have a direct impact on engineering self adaptive systems in general and the monitoring activity in particular. Examples of these include the time of change, the location of change, the frequency of change, etc. The description of these categories is presented in the next subsections.

**Temporal properties.** This category of change tries to address the question of *when* to evolve or make changes. It includes the time of change and frequency of change.

- **Time of change:** it concerns with when the desirable changes take place during the software lifecycle phases which include the compile-time, load-time and run-time. Based on this dimension, three types of changes can be identified and described as follows:

    o **Static:** such a change is performed at the source code level. As a result, the software needs to be recompiled for the changes to take effect.
    o **Load-time**: changes take place here prior to the loading of software components (classes for instance) to the memory for execution.
    o **Dynamic:** changes take place while the system is being executed.

    Static or compile time changes are inappropriate for true self adaptive systems since adopting this time of change requires introducing downtimes to the system which is usually not acceptable or cannot be afforded. On the other hand, load-time and dynamic (run-time) types are the appropriate time to make changes with a higher preference to the runtime.

- **Frequency of change:** it concerns with how often changes are carried out. Changes may be performed continuously, periodically or randomly. We believe that the system type can influence the change frequency. For example, the nature of self adaptive systems imposes unpredictable occurrences of the system change since such changes could occur at any point of time. In contrast, changes in traditional systems may be performed periodically in which the system is turned off for a period of time called the downtime. Most self adaptive systems cannot afford this downtime as stated earlier in the time of change dimension.

**Object of change.** This category of change tries to address the question of *where* changes should be applied. This category includes the artifact to be changed and the granularity of the change.

- **Artifact:** in static evolution, many software artifacts are subject to change. Such artifacts include the requirements, design, source code and documentation. However, dynamic evolution is concerned with the running software itself.

    Although static evolution can be done for self adaptive systems, the primary concern and real reason for the emergence of this kind of systems is to manage and regulate the core system at run-time.

- **Granularity:** it refers to the scale of the artifact to be changed which can be either coarse-grained or fine-grained. A coarse-grained artifact may take the form of *class* or package in the object oriented paradigm whereas the fine-grained one can take the form of variables, methods or statements inside a method. Changes (adaptations) in self adaptive systems are usually performed on these two scales depending on the system under consideration and/or the system attribute (e.g. performance) that the self adaptation capability has to address. Usually, these two scales are referred to as *parametric* and *structural* changes.

## 2.3 Active and Passive Monitors

The monitor component of the MAPE-K control loop can be operated in either an *active* or *passive* mode [14]. Active monitors report relevant data to interested parties only when some interesting event has occurred (e.g. database server connection failure). This is in contrast to the passive monitors where reporting data is performed at fixed delay (e.g. each 30 seconds) regardless of the occurrence of any event.

Both modes can bring some advantages and disadvantages. While an active monitor can exhibit some advantages by relieving the interested party (analyzer) from processing a large volume of data, the burden will be shifted to the monitor and sensor components which takes the form of an additional instrumentation of the managed system to track previous system states and detect some events that might indicate a change in the managed system or its environment. In contrast, in a passive mode, the monitor component is relieved from the additional instrumentation but the overhead here takes the form of continuous reporting of data to the interested party [15].

## 2.4 Instrumentation Techniques

The instrumentation in the software context is the mechanism of integrating the monitoring capabilities into the software system to be observed and managed. There are many approaches to enable instrumentation of software systems with the aim of collecting information about the behavior and states of these systems at run time. Such runtime information is crucial to test whether or not the system is operating in compliance with some predefined goals.

In this section, we review the different techniques of instrumentation. Instrumentation can be achieved by middleware interception and the extension of source code through the application of aspect-oriented programming (AOP) or byte Code instrumentation. The following is a brief description of these techniques.

**Middleware Interception**. Middleware systems are used to build distributed systems by abstraction the platform specific details away from the developer. Example of Middleware systems include COBRA and JEE. In [16], to mange distributed systems, a kind of instrumentation was applied which is referred to as *on demand distributed software instrumentation*. Also as in [17], the instrumentation code is promoted as a new middleware service.

**Aspect Oriented Programming (AOP).** Aspect-oriented programming languages aim to support the separation of crosscutting concerns. Cross-cutting concerns are concerns that affect the implementation of many modules in a system. Examples of such concerns include security, monitoring, logging, etc. AOP has been adopted by several researchers to support the implementation of many activities in self adaptive systems. In particular, AOP provides support for monitoring, event aggregation and dynamic adaptation [18]. AspectJ [19], an extension of Java programming language, is the most well known aspect-oriented programming model. Four fundamental concepts can be identified on which AspectJ is based. These concepts are *Aspect, joint point, pointcuts* and *advice* which are described as follows:

- *Aspect*: an aspect, like any normal Java class, contains a state (a number of fields) and methods. Its primary task is to intervenes in the control flow of some other classes by inserting some extra functionality at specific joint points.

- *Joint point*: a join point is a well-defined event in the execution of a program, such as the call to a method, the access to an object field, the execution of a constructor, or the throwing of an exception.

- *Pointcuts*: pointcuts are a way of referring to a set of join points.

- *Advice*: an advice specifies an operation that runs at any join point matched by the associated pointcut. Additionally, an advice defines whether it is executed before, after, or around the affected join points.

**Byte Code Instrumentation.** Bytecode instrumentation is a process where new functionality is added to a program by modifying the bytecode of a set of classes before they are loaded by the virtual machine. Java bytecode instrumentation is largely used for implementing dynamic program analysis tools and frameworks. Examples of well known frameworks and tools include ASM [20], Javassist [21] and BCEL [22]. When working at the bytecode level, the source code is not needed and thus is not changed. Support of bytecode instrumentation in Java resides in package java.lang.instrument which has been introduced since JDK 1.5.

## 3. SELF ADAPTIVE SYSTEMS MONITORING MODEL

Based on the discussion presented in the background section, we could identify a set of concepts and components related to the monitoring activity of self adaptive systems. We here attempt to design a standard monitoring model that is able to accommodate the different scenarios and cases. We divide our presentation of the model into a number of categories, namely the framework characteristics, the monitoring model requirements, the monitoring activity components and their interplay. Next subsections elaborate on these categories.

### 3.1 Monitoring Modeling Dimensions

The monitoring process is concerned with a set of dimensions taking the form of *why*, *what*, *when* and *how* questions. Some of these dimensions are borrowed from works conducted in [23]. The answers to these questions should pave the way for constructing a standard monitoring model. The 'why' question is related to the purpose of monitoring the underlying system. In [11], systems are managed to exhibit one (or all) of four QoS properties: *self-healing*, *self-protecting*, *self-optimizing* and *self-configuring*. What to monitor aspect is related to the measurement of the system and environment properties that are so important to the managing system and have to be monitored in order to keep the managed system at a desirable and acceptable state. The 'when' question is concerned with how frequently the monitoring activity is performed. Existing monitoring approaches adopt one of two ways for evaluating a set of interesting properties: (1) event-triggered, where the monitor is only invoked when the system changes from state to another, and (2) time-triggered, where the monitor periodically interrupts the system execution and reads its state. Thus, monitoring interesting properties can be conducted at fixed delay, in response to an event and/or on demand. How to get interesting properties is related to the way these properties are collected. Usually a set of sensors is used to make direct measurements of these properties. However, measurements can also be inferred from the system state or success or failure of an operation (e.g. unresponsive server). Based on these aspects, we can define the following requirements for the monitoring activity:

- Specify QoS property
- Specify system property for monitoring
- Specify environment property for monitoring

International Journal of Software Engineering & Applications (IJSEA), Vol.7, No.6, November 2016

- Specify monitoring mode
- Define system state
- Deploy sensors
- Define event
- Define threshold

Figure 1 shows the UML use case diagram for the above listed monitoring system requirements. As shown in the Figure, some requirements depends on the existence of other requirements. For instance, the realization of the use case ' Define system state' requires addressing the use case 'specify system property for monitoring' as well as the use case 'specify environment property for monitoring'. Similarly, the realization of the two latter use cases requires defining a threshold which is represented by the use case 'define threshold'.

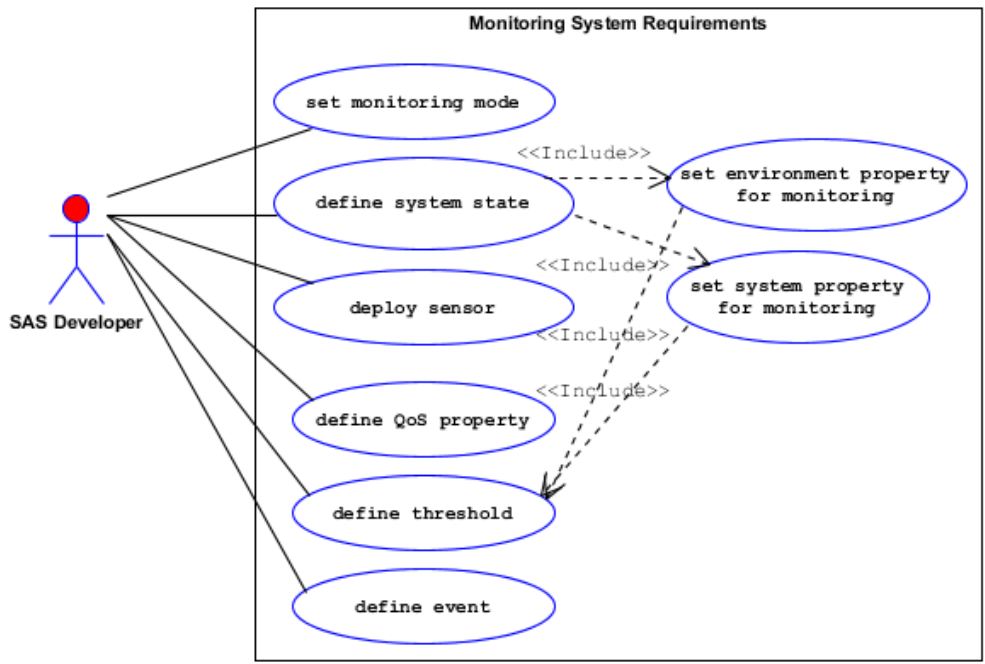

Fig. 1 Use case diagram for the monitoring activity requirements.

## 3.2 Monitoring Components

This category contains the core components of the monitoring activity which is extracted from the discussion of monitoring requirements presented in Section 3.2 . Recall that the task of monitoring is to monitor some interesting states of the software system under study via a set of sensors. Readings from sensors are aggregated and then logged into a repository which reflects the system state at different points in time. The obvious required components, some of which were presented in our previous work [24], identified here are: *monitoring controller*, *sensor*, *system state composer* and *logger*. With regard to the IBM blueprint [11], the monitoring controller and system state composer belong to the monitor component, the logger to the knowledge base and the sensor to the managed system touch points. Figure 2 depicts a high level interaction between these components.

*Monitoring controller*. It is responsible for taken care of collecting relevant data from the underlying system as well as deploying the necessary sensors to perform this task. This component is also in charge of tracing up the states (active/inactive) of the currently deployed sensors. Also, the construction of the system state composer is carried out by this component. Collected data from the deployed sensors, after being composed, is logged by the monitoring controller to the logger component located in the knowledge base as shown in Figure 1.

*Sensor*. Its sole responsibility is to collect data about the system property of high interest to the adaptation process and then send it to the monitor. Accomplishing this task can be conducted at fixed delay, in response to an event and/or on demand. When adopting the fixed delay form, the sensor sends the system property measurement to the monitor each t seconds where $t > 0$. However, the sensor may send the property measurement once the event of property threshold violation has occurred not waiting for the current time window to finish. Therefore, there are two kinds of sensor namely the time-triggered and event-triggered sensors. In the context of the implementation of the monitoring activity, this component contains the instrumentation code that is responsible for tracing up the system execution at specific points in the application.

*System state composer*. At runtime, the system state is represented by a combination of the values of system properties and the properties representing the environment or the context within which the system is operating. Each system has a desirable state driven by its goals and non functional requirements. Often the deviation from this desirable state is the trigger of the adaptation process.

*Threshold*. This is the value that the monitor component will compare against to decide whether the current value of the system or environment property is still within a desirable or acceptable range. A threshold might have two values for the lower and upper bounds. An example of a threshold would be if server CPU load becomes greater than 50%, or if load changes by more than 20%.

*System property*. This is the property that is of a direct connection and great interest to the adaptation process. This property is the target of the monitoring activity and the main concern of the monitor component is to keep its value within a desirable or acceptable range. Often, a threshold is used to accomplish this task. Examples of system properties include server load, server throughput, and response time and bandwidth usage. The system property contributes to the runtime system state.

*Environment property*. Tthe environment is defined as any external actor that affects the system in some way. Therefore, the environment property represents any contextual information that is external to the system in question and contributes to its runtime state. The currently connected clients in a client-server architecture represents an examples of such properties.

*System state*. At any point in time, the system in consideration is in one of possible states. Such states can describe the system structure, behavior or both. As stated earlier, the system state is composed of the values of the system and its environment properties at runtime.

*Event*. The event is something that occurs during the execution of the system under consideration which usually requires some actions to correct this undesirable situation. For instance, an out of stock product is one example of events in e-commerce applications.

*Logger*. This is the repository where collected data from the system and its environment is saved for later analysis.

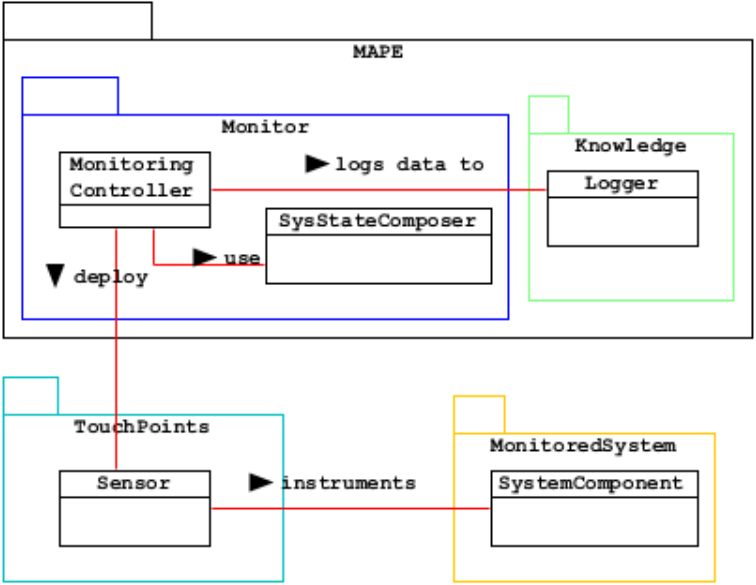

Fig. 2. Monitoring activity components.

## 4. CORE APPLICATION REQUIREMENTS

In this section, we introduce the requirement engineering of the core application (managed system) that is to be monitored by the monitor component. The managed system represents the system under development and is composed of a set of core functions. The following is the requirement that is related to the managed system and expected to be available in the framework:

- Provide interface for managed system model: This requirement is concerned with the ability of providing appropriate interfaces to define and model the managed systems. Fig. 3 depicts the UML use case diagram for the managed system requirement.

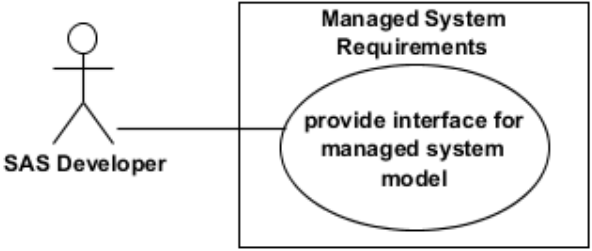

Fig. 3. UML use case diagram for the managed system requirement.

Some fundamental concepts of particular importance to this requirement are presented in [25] and described below:

*Domain*. The domain here is the system under consideration which comprises a number of tasks. Examples of domain include the healthcare , online banking, online travel agency and E-learning systems.

International Journal of Software Engineering & Applications (IJSEA), Vol.7, No.6, November 2016

*Task*. A task is a very high level goal that has to be addressed in order to address the overall system requirements. Each task, in turn, contains a set of services responsible for addressing and achieving that task. A task in a healthcare system is, for example, *monitor patient temperature.*

*Service*. A service is an abstraction of a software or hardware entity that has a role to play in addressing the task goal. These services, later at the code generation stage, are mapped into software components such as Web Services, CORBA, Java, .NET, etc. A temperature sensor or monitor is an example of service.

*Composite*. The services of a particular task coordinate with each other to address the purpose of that task. Such coordination, which involves a set of interactions, is encapsulated in an entity called composite. A composite might contain only one service. However, a useful composite is often composed of more than one service.

These concepts represent the process steps (in the order presented above) of the managed system requirements definition. In other words, the process starts with defining a *domain* and ends with constructing a *composite*. Then, each task and its associated services and *composite* will be described in the intention model, which is stored as an XML file. The latter is parsed to generate code in the form of Java classes by applying the cartridge of Java template. Also, the XML file is used to specify the properties, along with the threshold and events, to be monitored throughout the provisioning of an appropriate user interface. Fig. 4 shows a snapshot of a graphical user interface, developed in [26], for specifying interesting properties (parameters) for monitoring in the salt world domain which is an example of self organising systems [27].

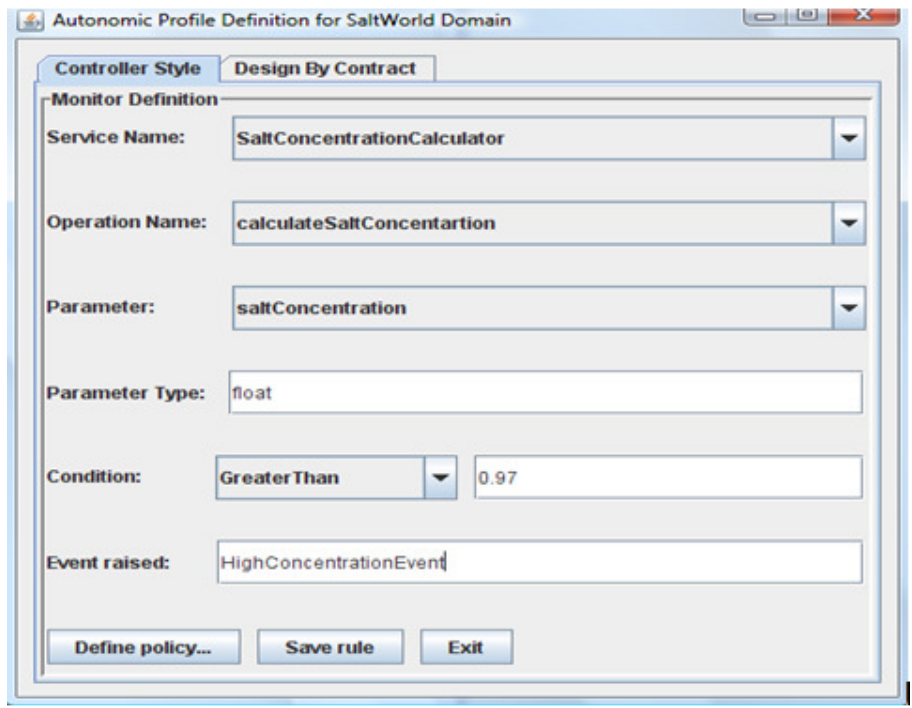

Fig. 4. System property specification for monitoring in Salt world domain.

## 5. FRAMEWORK IMPLEMENTATION

This section is dedicated to introduce the implementation part of the proposed monitoring framework. The framework is implemented in Java, so artifacts developed here take the form of packages, classes, interfaces, abstract classes, etc. These artifacts are distributed over a set of packages as follows:

- Monitoring package: this package contains classes and interfaces related to the monitoring activity that reside in the monitor component of MAPE-K.
-Sensing package: this package contains classes and interfaces related to the monitoring activity that reside in the managed system touch point component of MAPE-K.
-Knowledge package: this package contains classes and interfaces related to the monitoring activity that reside in the knowledge component of MAPE-K

**Monitoring package.** It is composed of the software components that carry out the fundamental role of the monitoring activity. All monitor controllers here implement an interface called `IMonitor` which is shown in Listing.1. As seen from the listing, the interface `IMonitor` defines the signatures of a set of methods that every monitor implementing this interface has to provide implementations for.

```
package monitoring;

public interface IMonitor {

    public void setMonitoringMode(int mode);
    public void setSystemState(SystemState ss);

}
```
Listing. 1. Definition of IMonitor Interface

A significant method of the interface is the method `setMonitoringMode` which enables the monitor to specify the monitoring mode which may take one of two values: *periodic* or *event-triggered*. The other defined method is `setSystemState` which is used by the monitor instance to set the system state. Recall the system state is composed of a number of properties ( for the system itself and its environment) of a particular interest to the monitoring activity of self adaptive systems.

The SystemState class is responsible for providing the necessary methods for performing the required tasks which are listed as follows:

- Adding a property to the system state
- Removing a property from the system state
- Retrieving a property (or all properties) from the system state
- 

Figure 3 depicts the UML class diagram for the `SystemState` class.

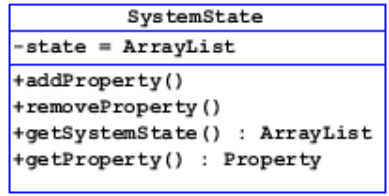

Fig. 5. UML class diagram for SystemState class.

Both the system and environment properties are abstracted in the `Property` class which is described with the UML class diagram shown in Fig. 4.

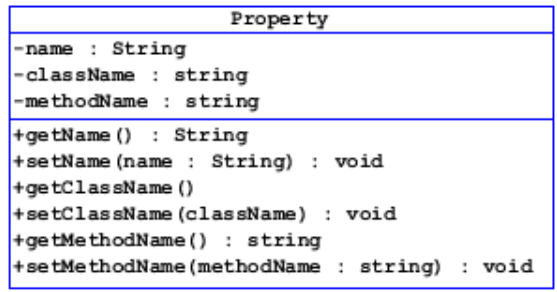

Fig. 6. UML Class diagram for Property class.

The `name`, `methodName` and `className` are three fields required to uniquely identify each system or environment property. Later, these are used by the monitoring controller to instrument the right component and watch for the change of the interesting property at the system runtime. Each property's value has to be maintained within a specific range which is often realized by specifying a threshold. Fig. 5 shows the UML class diagram of threshold class.

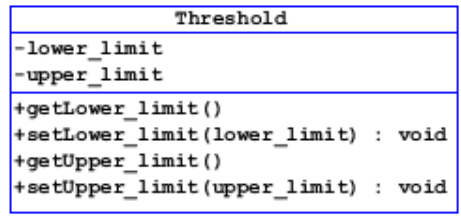

Fig. 7. UML class diagram of threshold class.

**Knowledge package.** This package contains the Java classes, interfaces, etc. that are responsible for the definition of the policy engine definition, logger and the like. However, in this framework, the focus will only be on the logger component since it is in a direct relation with the monitoring activity.

The system state is saved, periodically or upon undergoing some predefined change, into the *logger* component by the *monitoring controller*. The logger is used as the communication mechanism between the monitoring activity and the analysis activity. In other words, the change to the logger which is caused by new entries being logged by the monitor triggers the analysis activity.

International Journal of Software Engineering & Applications (IJSEA), Vol.7, No.6, November 2016

```
package knowledge;

import monitoring.SystemState;

public interface Logger {
    public void log(SystemState ss);
    public SystemState getData();
}
```

Listing. 2. Logger interface definition.

The data collection performed by the monitoring activity is assigned to the sensor which takes the form of instrumentation code. Instrumentation can be conducted in various ways; it can be done using AOP, BCI, etc. All instrumentation methods in our framework have to implements the `Instrumentation` interface which includes one single method called `instrument` that requires a parameter of `Property` type as illustrated in Fig. 6.

```
package sensing;

import monitoring.Property;

interface Instrumentation {
    public void instrument(Property p);
}
```

Listing. 3. Definition of Instrumentation interface.

Listing . 4 shows an implementation of the `Instrumentation` interface which adopts the AOP method.

```
package sensing;

import monitoring.Property;

class AspectBasedInstrumentation implements Instrumentation{

    Property property;

    @Override
    public void instrument(Property p) {
        // calling of instrumention logic code goes here
    }
}
```

Listing. 4. Java class of Aspect based Instrumentation.

As discussed in [24], the monitor plays the *observer* role in the Observer design pattern while the instrumentation or the sensor takes on the s*ubject* responsibility. Therefore, in addition to implementing the `IMonitor` interface, the monitor has to implement the *Observer* Interface as well and the sensor has to implement the `Subject` interface.

44

## 6. RELATED WORK

A considerable number of research papers has been published tackling the problem of runtime monitoring. A bytecode based instrumentation approach was proposed to monitor and control resource consumption in standard Java Virtual Machines (JVMs) [28], and to generate calling context-sensitive profiles for performance analysis [29, 30]. Dowson et al [31] discussed how to build monitoring into Java programs from the ground up with reflection technology to detect normal and exceptional system behavior. [32] proposed a runtime monitoring approach to ensure the adherence of the web services to the contracts specified between the involved parties. Also, active research has been conducted to reduce the monitoring overhead and propose adaptive monitoring. [33] presented approach to address the monitoring overhead by dividing the monitoring activity into two stages. In stage one, a set of metrics is chosen and monitored to detect any possible violations. Once abnormality is detected, stage two starts which requires extending the set of metrics selected in stage one to locate the defective system components. In [34], a different approach is taken to reduce the monitoring overhead. It is based mainly on decreasing the measurement frequency to reduce the overhead associated with the monitoring and increasing the frequency to provide higher assurance. In [35], an adaptive monitoring technique is introduced which is based on collecting system notifications from distributed nodes and adjusting the notification frequency according to the CPU load. Regarding the instrumentation techniques, many approaches have been proposed. A middleware based approach is adopted by [16, 17] to provide the required instrumentation facilities to the monitoring activity. In [20, 21, 22], the instrumentation process is based entirely on the manipulation of the bytecode and thus the instrumentation is performed prior to loading time of the target software component.

## 7. CONCLUSIONS

In this paper, we investigated the monitoring activity of the MAPE-K control loop in terms of the following aspects:

- The architecture of the monitoring activity which consists of the core components and the control and data flow between these components.
- The monitoring modes and their relevance to some specific situations.
- The appropriate and standard monitor interface with other activities of MAPE-K control loop (the analysis and Knowledge base).
- The adaptive monitoring and its importance to realizing cost effective (in terms of time, resource utilization, etc) self adaptive systems.

We also introduced a Java framework that includes the necessary software artifacts for enabling the monitoring activity in a standard, extendible and reusable way.

As future work and suggested research, the following issues need to be addressed:

- A more detailed description of the framework is needed in the form of an illustrative case study.
- Development of an appropriate graphical user interface to facilitate the definition of both the core application functions and the monitoring capabilities.
- Also, to include the user in the loop, a graphical user interface should be put in place to allow the system administrator to watch the system while it is being executed and may accept or reject some suggested actions.

## REFERENCES

[1] J. Camara, R. de Lemos, N. Laranjeiro, R. Ventura, M. Vieira. "Testing the robustness of controllers for self-adaptive systems". Journal of the Brazilian Computer Society 2014.

[2] P. Oreizy, M. M. Gorlick, R. N. Taylor et al., "An architecture-based approach to self-adaptive software". IEEE Intelligent Systems and Their Applications, vol. 14, no. 3, pp. 54–62, 1999.

[3] D. Iglesia, "A Formal Approach for Designing Distributed Self-Adaptive Systems" , PhD Thesis, Linnaeus University 2014.

[4] G. Brown, B. Cheng, H. Goldsby and J. Zhang. "Goal-oriented specification of adaptation requirements engineering in adaptive systems". In: ACM 2006 International Workshop on Self-Adaptation and Self-Managing Systems (SEAMS 2006), Shanghai, China, pp. 23–29, 2006.

[5] M. Ben Said, Y. Kacem, M. Kerboeuf, N. Ben Amor, M. Abid, "Design patterns for self-adaptive RTE systems specification". International Journal of Reconfigurable Computing, Vol. 2014 (2014).

[6] A. J. Ramirez and B. H. C. Cheng, "Design patterns for developing dynamically adaptive systems," in Proceedings of the ICSE Workshop on Software Engineering for Adaptive and Self-Managing Systems (SEAMS '10), pp. 49–58, ACM, USA, 2010.

[7] D. Garlan, S. Cheng , A. Huang, B. Schmerl , P. Steenkiste ."Rainbow: architecture-based self-adaptation with reusable infrastructure". IEEE Computor Society Press, vol. 37, no. 10, pp. 46–54, 2004.

[8] D. Garlan, S. Cheng and B. Schmerl."Increasing system dependability through architecture-based self-repair". In: de Lemos, R., Gacek, C., Romanovsky, A. (eds.) Architecting Dependable Systems. LNCS, vol. 2677. Springer, Heidelberg, 2003.

[9] U. Richter, M. Mnif, J. Branke, C. Muller-Schloer and H. Schmeck. "Towards a generic observer/controller architecture for organic computing". In: Hochberger, C., Liskowsky, R. (eds.) INFORMATIK 2006: Informatik f¨ur Menschen. GI-Edition – Lecture Notes in Informatics, vol. P-93, pp. 112–119 , 2006.

[10] C. Peper and D. Schneider."Component engineering for adaptive ad-hoc systems". In: ACM 2008 International Workshop on Software Engineering for Adaptive and Self-Managing Systems (SEAMS 2008).

[11] IBM. "An architectural blueprint for autonomic computing. IBM", 2005.

[12] R. Ali, A. Griggio, A. Franzen, F. Dalpiaz, and P. Giorgini. "Optimizing monitoring requirements in self-adaptive systems". In Enterprise, Business-Process and Information Systems Modeling, pp.362–377. Springer, 2012.

[13] J. Buckley, T. Mens, M. Zenger, A. Rashid and G. Kniesel. " Towards a taxonomy of software change". Journal of Software Maintenance and Evolution: Research and Practice, Vol. 5, No.17, pp.309-332, 2003.

[14] A. Mos and J. Murphy, "Compas: Adaptive performance monitoring of component-based systems", The Proceedings of the Second International Workshop on Remote Analysis and Measurement of Software Systems (RAMSS 04), pp. 35-40, 2004..

[15] A. J. Ramirez, B. H. C. Cheng and P. K. McKinley." Adaptive Monitoring of Software Requirements", Requirements@Run.Time (RE@RunTime) 2010 First International Workshop on, pp. 41-50.

[16] D. Reilly. "A Dynamic Middleware-based Instrumentation Framework to Assist the Understanding of Distributed Applications". PhD thesis, School of Computing and Mathematical Sciences. 2006, Liverpool John Moores University: Liverpool.

[17] W. Omar. "Self-Management Middleware Services For Autonomic Grid Computing". PhD thesis, School of Computing and Mathematical Sciences. 2003, Liverpool John Moores University: Liverpool.

[18] R. Haesevoets, E. Truyen, T. Holvoet and W. Joosen. "Weaving the Fabric of the Control Loop through Aspects". Proceedings of the First international conference on Self-organizing architectures, Cambridge, UK, September 2009.

[19] E. Hilsdale and J. Hugunin." Advice weaving in aspectj". In 3rd International Conference on Aspect-Oriented Software Development (AOSD), pages 26-35, ACM, 2004.

[20] http://asm.ow2.org/

[21] http://jboss-javassist.github.io/javassist/

[22] https://commons.apache.org/proper/commons-bcel/

[23] R. de Lemos et al., "Software Engineering for Self-Adaptive Systems: A Second Research Roadmap, Springer-Verlag Berlin Heidelberg 2013 (Eds.): Self-Adaptive Systems, LNCS 7475, pp. 1–32, 2013.

[24] Y. Abuseta, K. Swesi. "Design Patterns for Self Adaptive Systems Engineering". International Journal of Software Engineering & Applications (IJSEA), Vol.6, No.4, July 2015.

[25] Y. Abuseta, A.Taleb-Bendiab. "Model Driven Development based Framework for Autonomic Mobile Commerce Engineering", TAMoCo 2009: 51-60, 2009.

[26] Y. Abuseta. "AutoTaSC: Model Driven Development for Autonomic Software Engineering". PhD thesis, School of Computing and Mathematical Sciences. 2009, Liverpool John Moores University: Liverpool.

[27] M. Randles, H. Zhu, A. Taleb-Bendiab." A Formal Approach to the Engineering of Emergence and its Recurrence". in 2nd International Workshop on Engineering Emergence in Decentralised Autonomic Systems (EEDAS). 2007. Florida, USA: IEEE.

[28] W. Binder and J. Hulaas. "A portable CPU-management framework for Java". IEEE Internet Computing, vol.8, No.5, pp.74–83,. 2004.

[29] W. Binder. "A portable and customizable profiling framework for Java based on bytecode instruction counting". In Third Asian Symposium on Programming Languages and Systems (APLAS 2005), volume 3780 of Lecture Notes in Computer Science, pages 178–194,Tsukuba, Japan, Nov. 2005. Springer Verlag.

[30] W. Binder. "Portable and accurate sampling profiling for Java". Software: Practice and Experience, Vol. 36, No. 6, pp.615–650, 2006.

[31] D. Dawson, R. Desmarais, H. Kienle and H. Müller. " Monitoring in Adaptive Systems using Reflection", Proceedings of the 2008 international workshop on Software engineering for adaptive and self-managing systems, Pages 81-88.

[32] A. Lomuscio, W. Penczek, M. Solanki, and M. Szreter. "Runtime monitoring of contract regulated web services". In Proceedings of the 12th International Workshop on Concurrency, Specification and Programming (CS&P09), 2009.

[33] M. Munawar. " Adaptive Monitoring of Complex Software Systems using Management Metrics", PhD Thesis, University of Waterloo 2009.

[34] K. Clark, M. Warnier and F. Brazier. "Self-adaptive service monitoring", ICAIS'11 Proceedings of the Second international conference on Adaptive and intelligent systems, 2011.

[35] H. Keung, J. Dyson, S. Jarvis, G. Nudd. "Self-adaptive and self-optimising resource monitoring for dynamic grid environments". In: Database and Expert Systems Applications, 15th International Workshop on. pp. 689- 693 (Aug 2004).